\documentclass{article}

\usepackage{graphicx}
\usepackage{amsmath,amssymb}
\usepackage{color}

\begin{document}
\title{ \vspace{-2.5cm} \bf \large Thermodynamics at the microscale: from effective heating to the Brownian Carnot engine}
\author{L. Dinis$^1$, I.~A. Mart\'\i nez$^2$, \'E.~Rold\'an$^{3,4}$, J.~M.~R Parrondo$^1$, and R. A. Rica$^5$ }

\maketitle
\noindent{$^1$ GISC and Dpto.\ F\'\i sica At\'omica Molecular y Nuclear. Universidad Complutense de Madrid, Madrid, Spain}

\noindent{$^2$ Laboratoire de Physique, Ecole Normale Superieure, CNRS UMR5672 46 All\'ee
  d'Italie, 69364 Lyon, France.
}
\noindent{$^3$ Max Planck Institute for the Physics of Complex Systems, N\"othnitzer Str. 38, 01187
  Dresden, Germany.
}

\noindent{$^4$ Center for Advancing Electronics Dresden, cfaed, Germany.}

\noindent{$^5$ ICFO-Institut de Ciencies Fotoniques, The Barcelona Institute of Science and Technology, 08860 Castelldefels, Barcelona, Spain.}

\begin{abstract}
We {review a series of} experimental studies of the thermodynamics of nonequilibrium processes at the microscale. {In particular, in these experiments} we studied the fluctuations of the thermodynamic properties of a single optically-trapped microparticle immersed in water and in the presence of external random forces. In equilibrium, the fluctuations of the position of the particle can be described by an effective temperature that can be tuned up to thousands of Kelvins. {Isothermal and non-isothermal thermodynamic processes that also involve changes in a control parameter were implemented by controlling the effective temperature of the particle and the stiffness of the optical trap. Since truly adiabatic processes are unfeasible in colloidal systems, mean adiabatic protocols where no average heat is exchanged between the particle and the environment are discussed and implemented. By concatenating isothermal and adiabatic protocols, it is shown how a single-particle Carnot engine can be constructed. Finally, we provide an in-depth study of the fluctuations of the energetics and of the efficiency of the cycle.}
\end{abstract}
\noindent{Keywords: \it stochastic thermodynamics, Brownian motors, stochastic efficiency\/}
\section{Introduction }

Technological advances on micromanipulation and force-sensing techniques have brought access to  dynamics and energy changes in physical systems where thermal fluctuations are relevant~\cite{ashkin1970acceleration,VisscherNature1999,bustamante2005nonequilibrium,millen2014nanoscale}. 
At micro and nano scales, the combination of stochastic energetics~\cite{sekimoto2010stochastic} and fluctuation theorems~\cite{bochkov1981nonlinear,evans1993probability,jarzynski1997nonequilibrium,crooks1999entropy} have provided a robust theoretical framework that studies thermodynamics at small scales, thus establishing the emerging field of {\em stochastic thermodynamics}~\cite{seifert2012stochastic}. The study of the laws that govern the fluctuations of the energy transfers in nonequilibrium non-isothermal processes  are at the core of the key objectives of the theory of stochastic thermodynamics and the development of efficient artificial nanomachines. Until recently, the design of microscopic heat engines has been restricted to those cycles formed by isothermal processes or instantaneous temperature changes~\cite{blickle2012realization}.

In this paper we review our study of non-isothermal processes and thermodynamic cycles {realised with} a single single colloidal particle {as the working substance. The ultimate goal of this study was the implementation of a microscopic Carnot engine. In order to achieve this goal, we had to develop different experimental and theoretical tools, as we discuss below. These can be summarised in three achievements.} First, the ability to  accurately tune the effective temperature of the particle both under equilibrium~\cite{gomez2010steady,martinez2013effective,berut2014energy} and nonequilibrium driving~\cite{mestres2014realization}. Second, the possibility of measuring kinetic energy changes of a colloidal particle at low sampling rate~\cite{roldan2014measuring}, permitting a full description of the thermodynamics of the particle. Third, the realization of adiabatic process in the mesoscale by conservation of the full phase space entropy of the Brownian particle~\cite{martinez2014adiabatic}. 

Notably, the Carnot cycle is a fundamental building block of thermodynamics and is of paramount importance in the understanding of the Second Law of thermodynamics~\cite{carnot1872reflexions}. Sadi Carnot set a fundamental bound on the efficiency {that} a motor working between two thermal baths can achieve. Although it has attracted considerable attention from the theoretical point of view \cite{sekimoto200carnot,PhysRevE.90.042146, PhysRevE.85.011127}, the experimental realization of Carnot engine has been elusive to experimentalists because of the difficulty to design adiabatic protocols. Here, we {discuss} an experiment that is analogous to the Carnot cycle for a classical piston. The thermodynamic characterization of the engine shows universal features in the behavior of micro-engines. At the microscale, energy exchanges become fluctuating: the bound no longer applies and the macroscopic second law has to be replaced by fluctuation theorems. Using our setup we are able to verify recent theoretical prediction of statistical properties of the efficiency of stochastic engines derived using fluctuation theorems~\cite{verley2014unlikely,verley2014universal,gingrich2014efficiency}.

This paper is organised as follows: Section~\ref{sec_experimental} describes the main features of our experimental setup. Section~\ref{sec_measuring} shows how to measure the instantaneous velocity of Brownian particles at low sampling rates. Section~\ref{sec_microadiabaticity} is devoted to the description of microadiabatic processes. The construction of a Brownian Carnot engine is discussed in Sec.~\ref{sec_brownian} and Sec.~\ref{sec_open_problems} introduces the concluding remarks and discusses open problems.

\section{Experimental setup \label{sec_experimental}}

\begin{figure}[h]
\centering
\includegraphics[width=11cm]{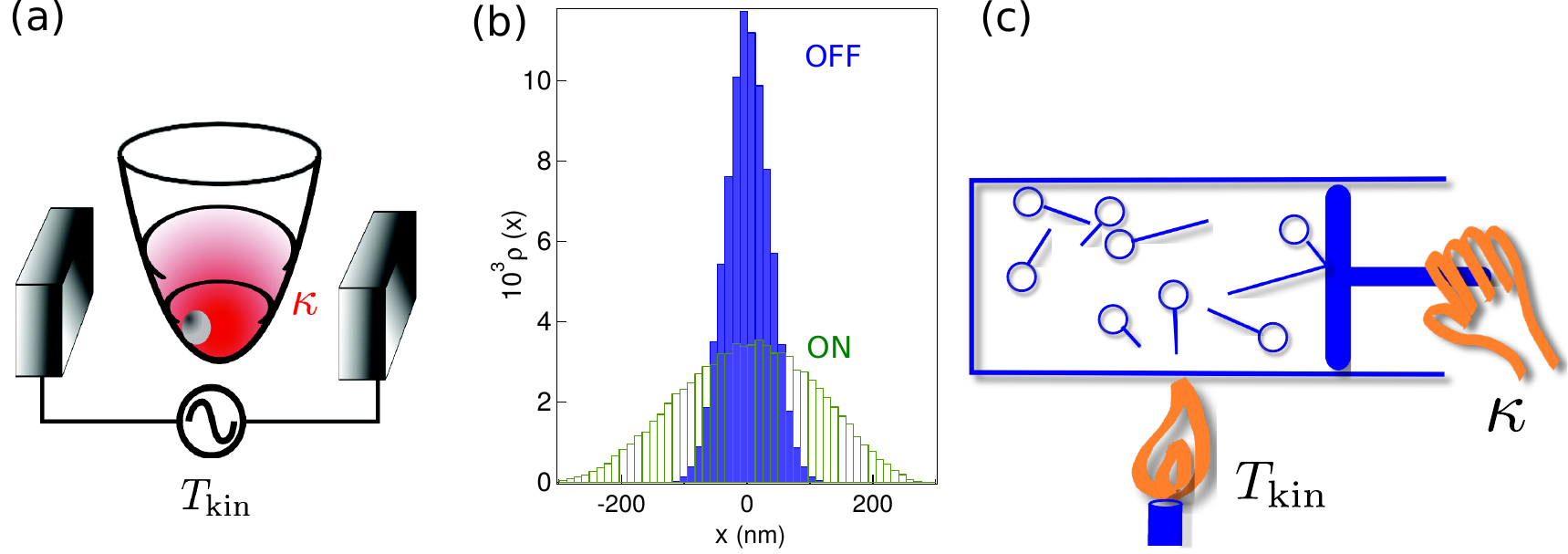}
\caption{\label{fig_setup} \textbf{Schematic of the experimental setup}. (a) A colloidal particle is trapped with an infrared laser, which creates a parabolic potential of stiffness $\kappa$. Two electrodes in the chamber induce a randomly fluctuating electric field that simulates the ``kicks'' of a higher temperature reservoir. { (b) Histograms of particle position with (marked ``ON'', green) and without (``OFF'', blue) the additional electric field. (c) By modifying trap stiffness and temperature, our system transforms heat into work in analogy with an ideal gas in a piston.} }
\end{figure}

\begin{figure}[h]
\centering
\includegraphics[width=12cm]{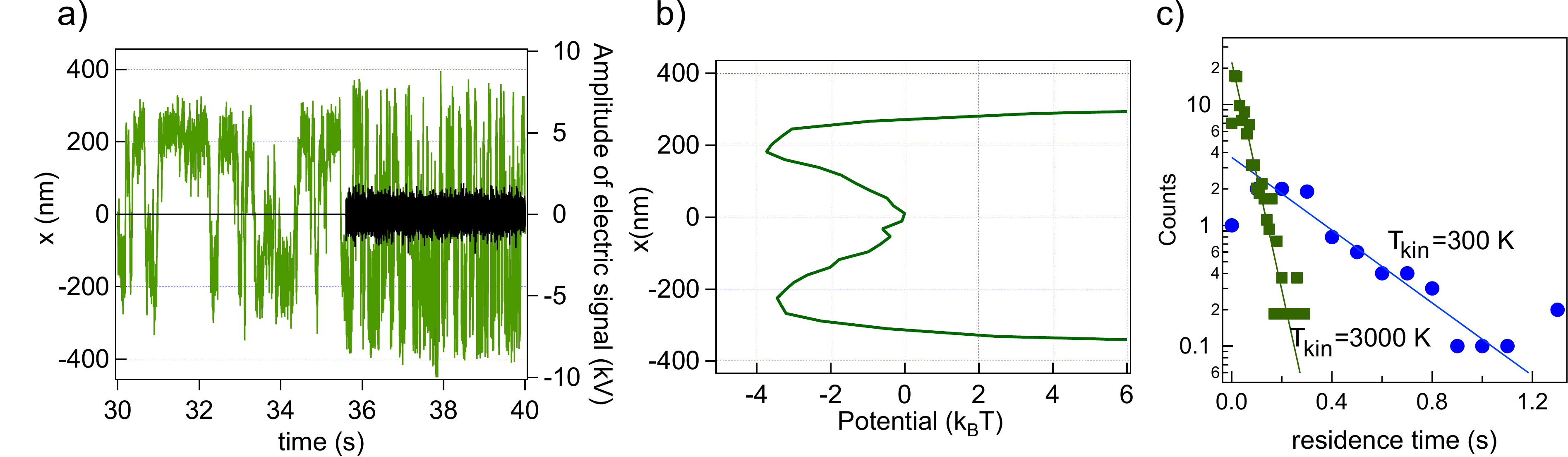}
\caption{\label{fig_Kramers} \textbf{Changing the temperature to change the mean life of a metastable state~\cite{martinez2013effective}}. First experimental evidence of the change of the characteristic times in a double well potential using an effective temperature  (a) The position traces (green lines, left axis) of the sphere in the double-well potential. The black lines (right axis) show the voltage on the electrodes. The additional noisy signal was  switched  on  at  35.5  s.  As  can  be  seen,  the  frequency  of  the Kramers transitions increases. (b) The trapping potential obtained at room temperature. (c) Probability of the residence time of the Kramers transitions (in counts, log10 scale) at room temperature (green squares) and at 3000 K (blue circles). }
\end{figure}

{Our experimental setup is sketched in figure~\ref{fig_setup} and consists of a single polystyrene bead of radius $R=500$ nm immersed in water which is managed by the optical tweezers technique\cite{martinez2013effective}. The optical potential is created by a highly focused infrared laser that creates a quadratic potential around its focus, $U(x)=\frac{1}{2}\kappa x^2$, where $x$ is the position of the particle with respect to the trap center and $\kappa$ the stiffness of the trap. Two aluminium electrodes located in the ends of a custom-made chamber are used to apply a voltage of controllable amplitude along the $x$ axis.} A random signal generated with a random number generator is fed to the electrodes which results in a random force exerted to  the colloidal particle. Under the external random force, the amplitude of the fluctuations of the position of the particle increase (see figure \ref{fig_setup}b). {We demonstrate that such enhancement of the fluctuations can be interpreted as an increase of the effective or kinetic temperature of the particle}~\cite{martinez2013effective,mestres2014realization,dieterich2015single}. A similar technique has recently been used to simulate the hot bath in a heat engine constructed with a single atom~\cite{2015ScienceRosnagel}. In our setup,  the effective temperature of the particle is related to the amplitude $\sigma$ of the generated random force~\cite{martinez2013effective}:
\begin{equation}
  T_{kin}=T_{w}+\frac{\sigma^2}{2k\gamma}.
\end{equation}
where $T_w$ is the actual temperature of the water the colloidal particle is immersed in, $\gamma$ the friction coefficient and $k$ Boltzmann's constant. In steady state, it can be directly obtained from the enhanced position fluctuations. We first define an effective temperature for the particle obtained from the position fluctuations
\begin{equation}
  T_\mathrm{part}\equiv\frac{\kappa\langle x^2\rangle}{k}\quad.
  \label{eq:Teff}
\end{equation}
In the steady state, equipartition theorem implies that $\langle U(x)\rangle = \frac{1}{2}kT_{\rm kin}$, where $\langle \cdot\rangle$ denotes an average over many realizations, which here coincides with a steady state average. Since $\langle U(x)\rangle=\frac{1}{2}\kappa\langle x^2\rangle$, the kinetic temperature is proportional to the variance of the position and $T_\mathrm{part}=T_\mathrm{kin}$. Notice that out of equilibrium however, both temperatures may differ. 
{A similar analysis can be done studying the power spectra density of the equilibrium trajectories.} For simplicity, we will denote in the following the effective or kinetic temperature as  $T$. With a suitable amplitude of the external random force, the effective temperature of the particle can be raised even beyond the vaporisation temperature of the water {The highest temperature we reached so far is $\sim6000\,$K~\cite{roldan2014measuring}, but there are no fundamental limits above it}. 
{This increase of the thermal energy of the system can be used broadly, from the excitation of metastable states (see figure~\ref{fig_Kramers} and \cite{dieterich2015single}) in biophysics to our present study: a fundamental study of the fundamental laws of thermodynamics in the mesoscopic scale.}

{The stiffness of an optical trap is proportional to the laser power \cite{mazolli2003theory}}. Therefore, we can easily control $\kappa$ by tuning the intensity of the laser.  Our setup can be therefore conceived as a microscopic machine able to implement any thermodynamic process in which the stiffness of the trap and the effective temperature of the particle can change with time arbitrarily following a protocol $\{T(t),\kappa(t)\}$. Under nonequilibrium driving, the effective temperature defined in (\ref{eq:Teff}) and that measured from nonequilibrium work fluctuations do coincide~\cite{mestres2014realization}. {Consequently, our device can be used in the study of stochastic thermodynamic properties even far beyond equilibrium.}


\section{Measuring kinetic energy changes of a Brownian particle from low acquisition rates \label{sec_measuring}}

When the effective temperature of the particle is changed, both the position and the velocity degrees of freedom, and therefore the potential and kinetic energy of the particle do change in time. Even in {the} quasistatic limit, the fluctuations of the velocity of the particle are affected by the change of the temperature by virtue of equipartition theorem, $\frac{1}{2}m \langle v^2\rangle=\frac{1}{2}kT_{\rm kin}$, with $m$ the mass of the particle. The experimental access {to} the instantaneous velocity of Brownian particles is therefore needed to measure the fluctuations of the kinetic energy at the mesoscale. The energy exchange between the surrounding bath and the particle velocity  due to the temperature change  affects both the energetics of the particle~\cite{sekimoto2010stochastic, schmiedl2008efficiency} but also its total entropy~\cite{seifert2005entropy,Dunkel2005time}.
 {Therefore,} the contribution of kinetic energy cannot be neglected in any thermodynamic process that involves temporal~\cite{schmiedl2008efficiency} or spatial~\cite{bo2013entropic} temperature changes. The underdamped Langevin equation~\cite{Langevin1908} provides an accurate description of the dynamics of microscopic particles under temperature changes:

 {
 \begin{equation}
   m\ddot x(t)=-\gamma \dot x(t)-\kappa x(t) +\xi(t)+\zeta(t)
      \label{eq_langevin}
 \end{equation}
 where $x$ represents the position of the particle with respect to the center of the trap of stiffness $\kappa$, $\xi(t)$ the random force exerted by the thermal bath, and $\zeta(t)$ is the additional electric random force used to mimic a higher temperature reservoir, both white Gaussian noises. There are two characteristic time scales in the dynamics of a particle following (\ref{eq_langevin}). On the one hand, the position of the particle tends to relax to the center of the trap in times of the order of $\gamma/\kappa$, which is of the order of the millisecond for the colloidal particles used in our experiments.}

On the other hand, due to the erratic motion caused by collisions with the fluid {molecules}, the colloidal particle returns momentum to the fluid at times $\sim m/\gamma$, 
thus defining an inertial characteristic frequency $f_p=\gamma/2\pi m$. The momentum relaxation time is of the order of nanoseconds for the case of microscopic dielectric beads immersed in water. Therefore, in order to accurately measure the instantaneous velocity of a Brownian particle, it might be necessary to sample the position of the particle with sub-nanometer precision and at a sampling rate above MHz~\cite{li2010measurement}. 

\begin{figure}
\centering
\includegraphics[width=0.8\linewidth]{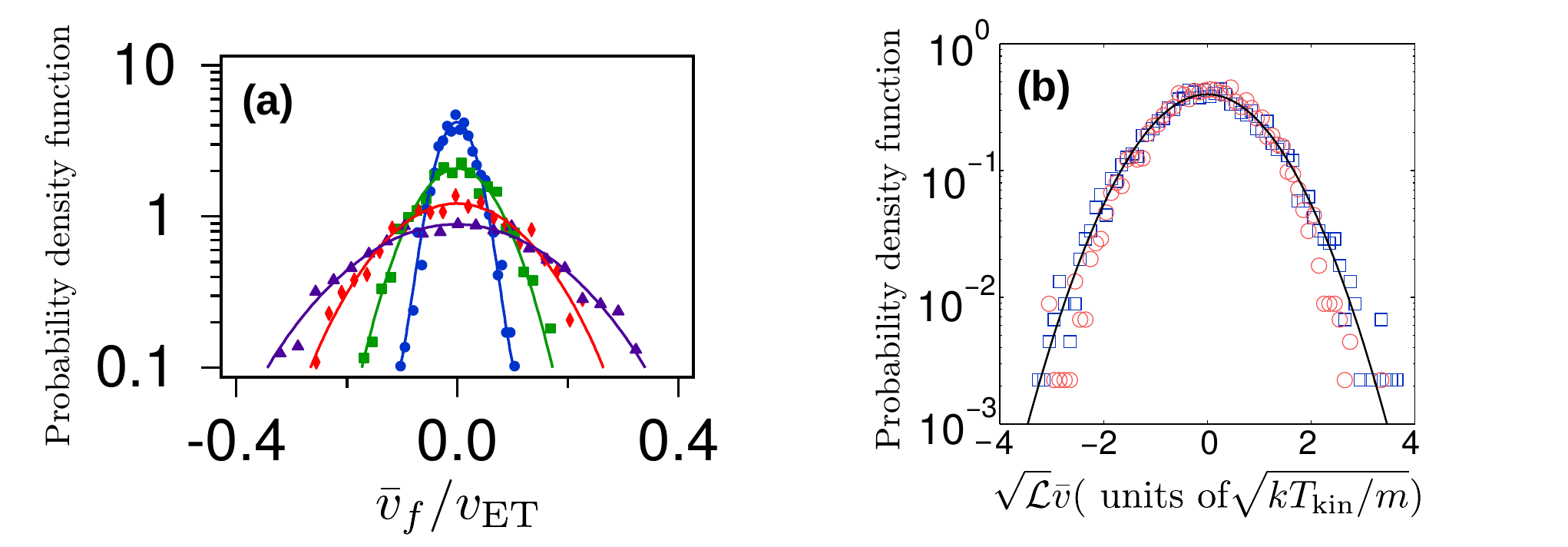}
\caption{\textbf{Time-averaged velocity distributions and reconstruction of the instantaneous velocity \cite{roldan2014measuring,martinez2014adiabatic}:} a) Probability density function of the time averaged velocity normalized to the standard deviation predicted from equipartition $v_{\rm ET} = \sqrt {kT/m}$. Different curves are obtained for the same stiffness $\kappa=(18.0 \pm 0.2)\,\rm pN/\mu m$ but for increasing values of the amplitude of the electric noise, corresponding to the following effective temperatures: $T=300\, \rm K$ (blue circles), $1060\, \rm K$ (green squares), $3100\, \rm K$ (red diamonds), $5680\, \rm K$ (magenta triangles). Each curve is obtained from a single measurement of $\tau=12\rm s$, sampled at $f = 5\, \rm kHz$. Solid lines are fits to Gaussian distributions. {Although the width of the distributions increases with effective temperature, it is several orders of magnitude smaller than expected for the instantaneous velocity. } b) Probability density function obtained by rescaling the time averaged velocity by the factor $\sqrt{L}$ given by the square root of (\ref{eq:FactorBdeC_trapped}) at temperature  $T=300\, \rm K$ (blue squares) and $T=600\, \rm K$ (red circles). After rescaling, the velocity distributions are described by Maxwell distribution (black curve)  \label{fig_velocity_pdf}}
\end{figure}

 In our experiment, we do not have direct access to the instantaneous velocity of the trapped bead due to the limited sampling speed of at most $10$kHz. However, even at this limited sampling frequencies, the effect of varying the effective temperature  of the particle can be appreciated in the velocity distribution. In fact, the  distributions are wider for higher temperature but the total width of the distributions is much smaller than expected from the equilibrium equipartition theorem, as shown in figure~\ref{fig_velocity_pdf}.
This slight dependence of the velocity distributions with temperature indicates that there is some information about the temperature that can be extracted from the trajectories even at low sampling rates. Applying the technique described in \cite{roldan2014measuring} we extrapolate the instantaneous velocity from the time averaged velocity  $\overline v_f$ obtained sampling the position at a frequency $f$ in time intervals $\Delta t = 1/f$. In an equilibrium or quasistatic processes, the mean squared time-average velocity and the mean squared instantaneous velocity at time $t$ of a particle trapped with a harmonic trap are related by
\begin{equation}
  \langle v^2(t)\rangle=\frac{ \langle \overline{v}^2_f(t)\rangle}{L(t)}\quad.
  \label{eq:velBDC}
\end{equation}
The factor $L(t)$ in (\ref{eq:velBDC}) is given by
\begin{equation}
  L(t)= 2f^2\left[ \frac{1}{f_0^2} + \frac{e^{-\frac{f_p}{2f}}}{f_1} \left(  \frac{e^{-f_1/f}}{f_p+2f_1} - \frac{e^{f_1/f}}{f_p-2f_1}  \right)\right]\quad,
  \label{eq:FactorBdeC_trapped}
\end{equation}
and depends on the sampling rate $f$, the momentum relaxation frequency $f_p$,  $f_0=\sqrt{f_p f_\kappa}$, $f_\kappa=\kappa(t)/2\pi\gamma$ and $f_1=\sqrt{f_p^2/4-f_0^2}$.

\begin{figure}
   \centering
   \includegraphics[width=8cm]{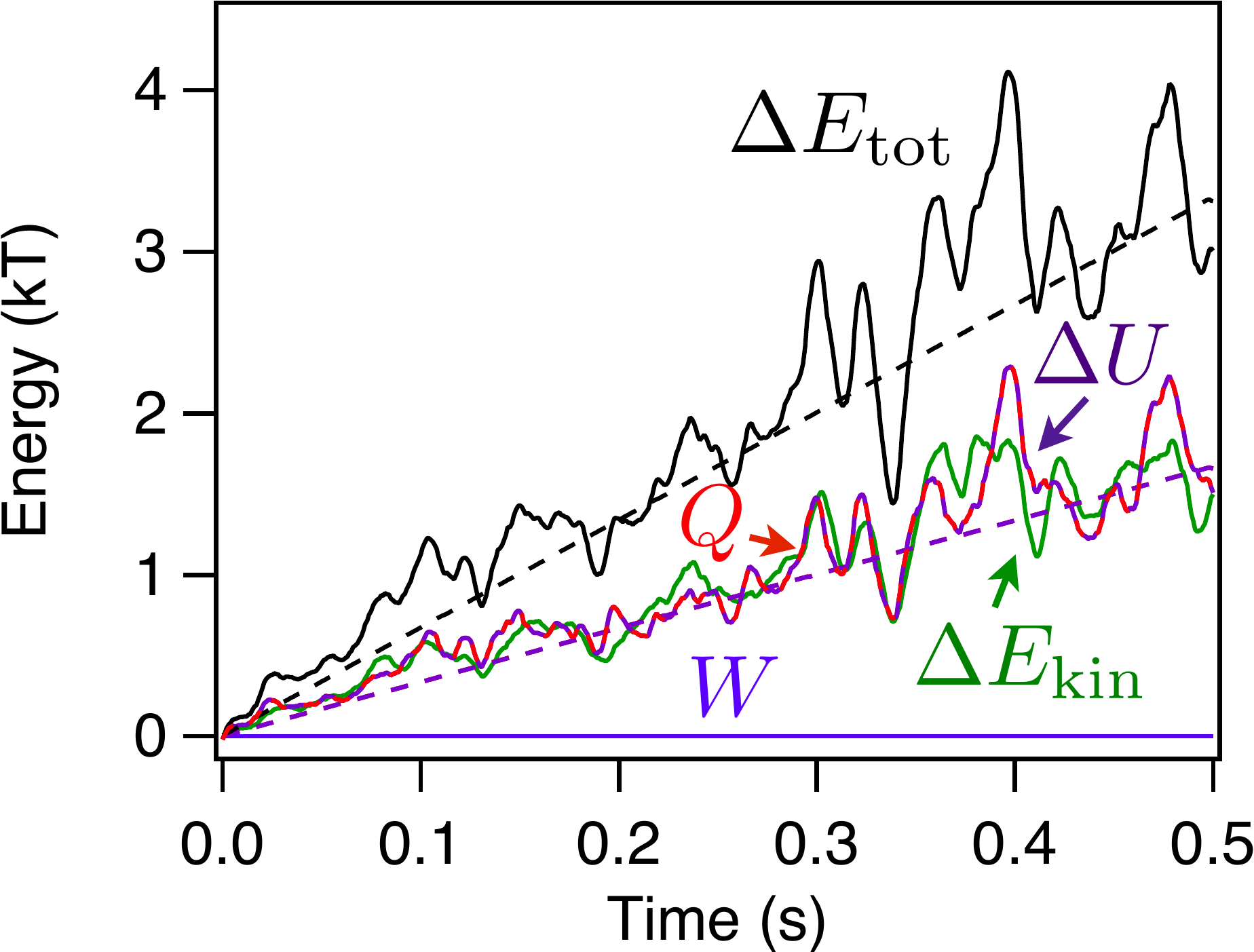} 
   \caption{\textbf{Energetics of a non-isothermal process \cite{martinez2014adiabatic}:}Experimental ensemble averages of the cumulative sums of thermodynamic quantities as functions of time in a non-isothermal process, where $T$ changes linearly with time from $300~\rm K$ to $1300~\rm K$ at constant $\kappa=(18.0 \pm 0.2)\,\rm pN/\mu m$:  Work (blue solid line), heat (red solid line), potential energy (magenta solid line), kinetic energy (green solid line) and total energy (black solid line).  Potential energy and heat are overlapped in the figure. Theoretical predictions are shown in dashed lines. All the thermodynamic quantities are measured from low-frequency sampled trajectories of the position of the particle, with $f=1\,\rm kHz$. Other parameters: $\tau=0.5\,\rm s$; ensembles obtained over $900$ repetitions.}
   \label{fig:IC_energetics}
 \end{figure}

Then, the average kinetic energy change along a protocol can be computed from measurements of the time averaged velocity sampled at a given frequency $f<f_p$ using
\begin{equation}
  \langle\Delta E_{\rm kin} (t)\rangle = \frac{1}{2}m\,\Delta\langle v^2(t)\rangle= \frac{1}{2} m\, \Delta\left(\frac{\langle \overline{v}^2_f(t)\rangle}{ L(t)}\right)\quad.
		\label{eq:EkinTAV}
\end{equation}
Equation (\ref{eq:EkinTAV}) allows to track kinetic energy changes in \emph{non-isothermal} processes from measurements of finite time averaged velocities. In figure~\ref{fig:IC_energetics}, we show the experimental and theoretical values of the ensemble averages of {some} thermodynamic quantities for a process in which the stiffness of the trap is held fixed and the effective temperature of the particle is changed linearly in time. In the figure, heat and work are obtained from $1 \rm kHz$ sampling of the particle position according to their definitions in stochastic thermodynamics~\cite{sekimoto2010stochastic} and the potential energy change as $\Delta U = Q+W$. The average kinetic energy is extrapolated from the time averaged velocity using (\ref{eq:EkinTAV}) and the total energy is obtained as the sum $\Delta E_{\rm tot} = \Delta U + \Delta E_{\rm kin}$. Since the control parameter (stiffness) does not change in this example, there is no work done on the particle along the process and $\Delta U=Q$. 
Temperature changes slowly {enough} so that the whole process is quasistatic and instantaneous values satisfy equipartition along the process (see \ref{sec_appendix_quasistatic}).
For instance, the potential energy change is given by $\langle \Delta U(t)\rangle = \langle Q(t) \rangle = \frac{k}{2}[T(t)-T(0)]$. Our measurement of kinetic energy is in accordance with equipartition theorem as well, yielding $\langle \Delta E_{\rm kin} (t) \rangle = \frac{k}{2}[T(t)-T(0)]$. Adding the kinetic and internal energies, we recover the expected value of the total energy change of the particle, $\langle \Delta E_{\rm tot} (t) \rangle = k [T(t)-T(0)]$. As a result, combining our experimental technique with the extrapolation method to ascertain the velocity of the particle, we can {evaluate} the fluctuations of the thermodynamic quantities of non-isothermal thermodynamic processes at the microscale.

 \section{Microadiabaticity \label{sec_microadiabaticity}}

Among all the possible thermodynamic processes, we focus our attention in adiabatic protocols since they are essential building blocks of any Carnot cycle. In the classical definition, an adiabatic process is such that no heat is transferred between the system and the thermal bath. The classical notion of adiabaticity involves several practical difficulties when considering a microscopic colloidal particle immersed in water, where the heat exchanges between the particle and the bath are stochastic and unavoidable. Achieving adiabatic processes in the microscale is therefore a challenging task, that can be circumvented by designing mean adiabatic or {\em microadiabatic} protocols where no net heat transfers occur between the particle and the environment when averaging over many realizations of the same protocol.  Motivated by theoretical proposals~\cite{sekimoto200carnot,bo2013entropic}, we  show how to realize experimentally microadiabatic protocols where the full available phase-space volume of the particle and therefore the average heat transfer are maintained equal to zero along the protocol~\cite{martinez2014adiabatic}.

In the microscale, heat and entropy are stochastic quantities that vary between different realizations of the same process. In quasistatic thermodynamic processes, the control parameter changes more slowly than any relaxation time {(both the position and velocity relaxation time)} and the total entropy change vanishes when averaging over many realizations of the same process, 
\begin{equation}
 \Delta S_{\rm tot}(t)= \Delta S(t) +  \Delta S_{\rm env}(t)=0\quad,
\label{eq:sbalance}
\end{equation}
where $\Delta$ stands for changes with respect to the initial value and all the quantities in~(\ref{eq:sbalance}) are ensemble averages. Here $\Delta S(t) $ is the average system entropy change, which is given by the change of the Shannon entropy of the phase space density~\cite{seifert2005entropy}. The second term in~(\ref{eq:sbalance}) is the average environment entropy change,
\begin{equation}
 \Delta S_{\rm env}(t) = -\left\langle\int_0^t \frac{\delta Q(s)}{T(s)}\right \rangle\quad,
\end{equation}
where $\delta Q(t)$ is the stochastic heat exchange in $[t,t+dt]$ between the particle and the environment~\cite{SekimotoPTPS1998}. We define a {\em microadiabatic} protocol as a quasistatic process where the mean (ensemble average) system entropy remains constant in time, and therefore $ \Delta S(t) = 0$ for all $t$. In a microadiabatic protocol, $\langle \delta Q(t)\rangle = T(t)\, dS(t) = 0$.

Given the fast momentum relaxation, it is tempting to drop the velocity degree of freedom and restrict to an overdamped description.  Schmiedl and Seifert showed that, in the overdamped limit, the system entropy can be conserved keeping the position distribution constant which can be met by changing the temperature and the stiffness of the trap such that their ratio $T(t)/\kappa(t)$ is maintained constant in time~\cite{schmiedl2008efficiency}. {However,} temperature changes involve changes in the distribution of the velocity of the particle and the Schmiedl--Seifert  protocol is therefore {\em pseudo-adiabatic} in the sense that energy leaks between the particle and the environment {(heat) do} occur through the velocity degree of freedom. {Therefore, even if} the overdamped description gives an accurate approximation of the dynamics of a colloidal particle at the time resolution of our experiment, {the energetics is incomplete. In} order to achieve actual adiabatic processes,  a full underdamped description of the dynamics is mandatory~\cite{martinez2014adiabatic}.

In our experiments, we implement thermodynamic processes where both the stiffness of the trap and the temperature of the particle can be switched in time, $\Lambda(t)=\{T(t),\kappa(t)\}$. In a quasistatic thermodynamic process, the control parameter changes slower than any relaxation time of the system and the phase space density of the system can be described by Gibbs distribution throughout the whole process. In a harmonic optical potential, the Hamiltonian of the particle is $\mathcal{H}(x,v)=\frac{1}{2}\kappa x^2+\frac{1}{2}m v^2$. At a given time $t$, the state of the system is described by the canonical distribution $\rho(x,v;t)=\exp[-\beta(t) \mathcal{H}(x,v;t)]/Z(t)$, where $\beta(t)=1/kT(t)$, and $Z(t)$ is the partition function at time $t$, which is equal to
\begin{equation}
  Z(t)= \int \, dx\, dv\, e^{-\beta(t)\mathcal{H}(x,v;t)}=  \left( \frac{4\pi^2 k^2}{m}\right)^{1/2}  \left(\frac{T^2(t)}{\kappa(t)} \right)^{1/2}\quad.
\end{equation}
At time $t$, the free energy of the particle equals to
\begin{equation}
  F(t)=-kT(t) \ln Z(t) = -\frac{kT(t)}{2} \ln \left( \frac{4\pi^2 k^2}{m} \frac{T^2(t)}{\kappa(t)}   \right)\quad.
\end{equation}
The entropy of the particle at time $t$, $S(t)$, can be determined from its free energy, $S(t)=-(\partial F(t) / \partial T(t))$. As a result, 
the entropy change of the particle from $t=0$ to $t>0$, $\Delta{S}(t) = S(t)-S(0)$, equals to
\begin{equation}
  \Delta{S} (t) = k \Delta \left( \ln \frac{T^2(t)}{\kappa(t)} \right)\quad.
  \label{eq:DSUD}
\end{equation}
In a quasistatic thermodynamic protocol where the ratio $T^2(t)/\kappa(t)$ {\em does not} change in time, the entropy of the particle remains constant. Since in the quasistatic limit, $\langle Q(t)\rangle = T(t) \Delta S(t)$, this defines a {\em microadiabatic} process where the ensemble average of the heat transfer to the particle vanishes. 

\begin{figure}
\centering
\includegraphics[width=8cm]{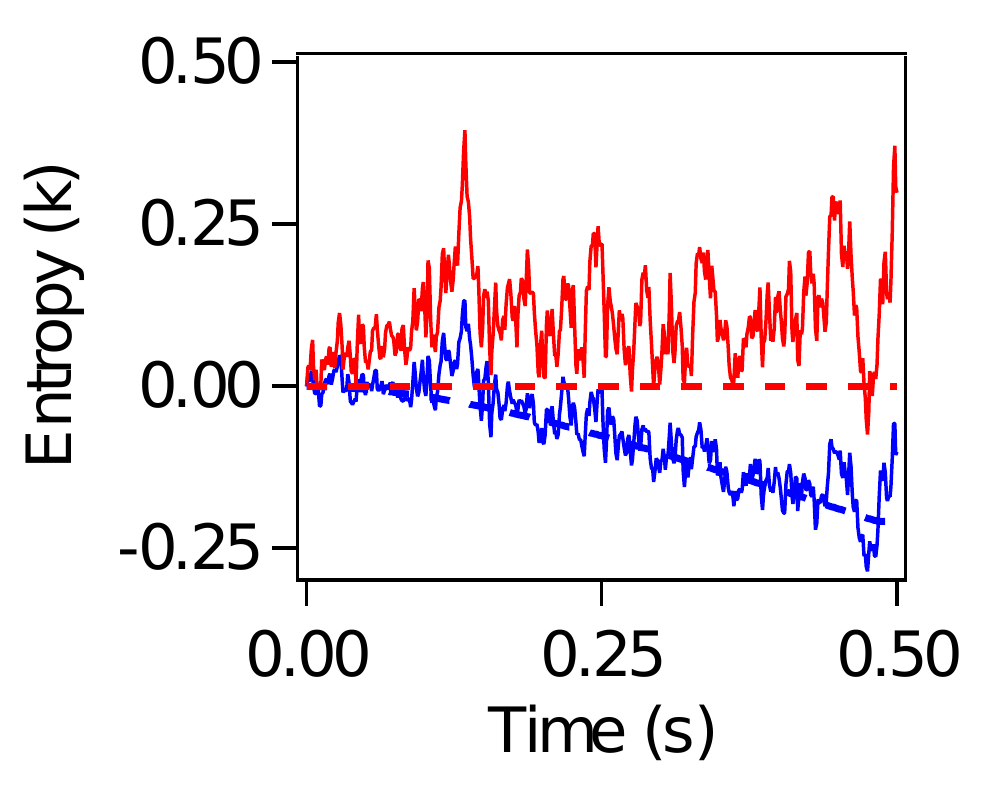}
\caption{\label{fig_entropy_adiabatic} {\textbf{Entropy change in adiabatic processes~\cite{martinez2014adiabatic}.} System entropy change in a adiabatic process where $T$ and $\kappa$ increase. The increase of entropy in the velocity degree of freedom (not shown) is approximately compensated by the positional degree of freedom $\langle \Delta S_x(t)\rangle $ (blue) to give a close to zero total entropy change $\langle \Delta S(t)\rangle$ (red). }}
\end{figure}

We then implement experimentally a microadabatic {protocol following this proposal}. The experimental protocol {has a duration} of duration $\tau = 0.5\,$s, which exceeds by several orders of magnitude the relaxation time of the particle in the trap, which is of the order of milliseconds~\cite{roldan2014measuring}. The driving is therefore slow enough to be considered as quasistatic. Figure \ref{fig_entropy_adiabatic} shows that the average system entropy ---and therefore also the average heat exchanged between the particle and the thermal bath (see adiabatic steps in figure \ref{fig_Cumulative_work_and_heat})--- vanishes in the microadabatic protocol, as predicted by the theory.

\begin{figure}
  \centering
  \includegraphics[width=8cm]{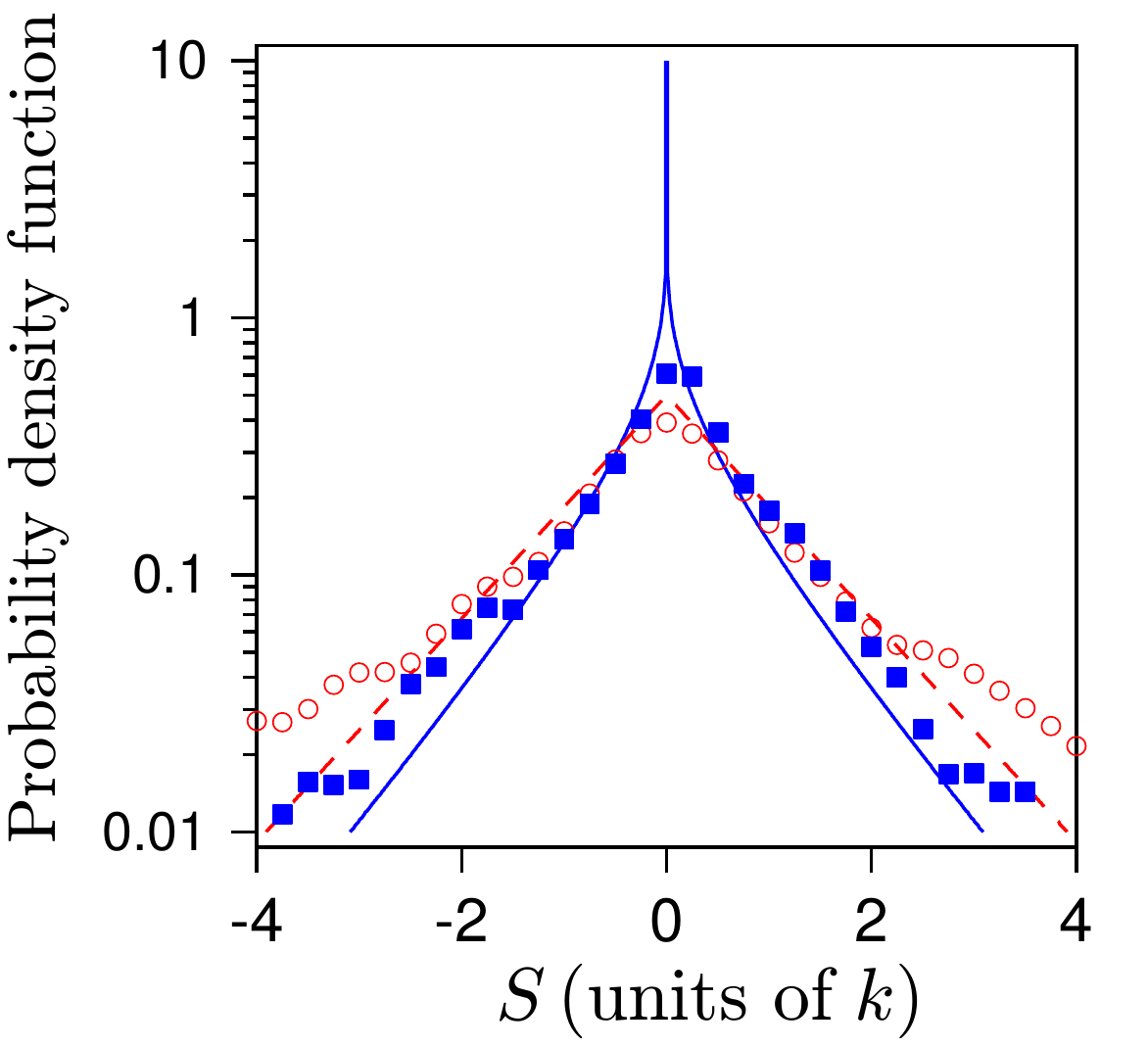}
  \caption{\label{fig_Fig5} \textbf{Statistics of system entropy change \cite{martinez2014adiabatic}:} Distribution of the system entropy change  $\Delta S_{x}$ in the overdamped description in a pseudo-adiabatic ($T/\kappa=\mathrm{const.}$) process (blue squares) and of the total system entropy change $\Delta S$  in a true adiabatic ($T^2/\kappa=\mathrm{const}$) process (red circles). The distributions are obtained from $900$ cycles. Theoretical distributions for $\Delta \mathcal{S}_x$ (blue solid curve) and  $\Delta \mathcal{S}$ (red dashed curve)  are also shown. }
\end{figure}

The stochasticity of the entropy change is analyzed in figure~\ref{fig_Fig5}, which shows the distribution of the entropy change in different realizations of the microadiabatic process. The distribution of the stochastic system entropy $\Delta s$ is exponential and symmetric around zero, {(see Supplementary information in \cite{martinez2014adiabatic}):}
\begin{equation}
\rho(\Delta s) = \frac{1}{2k} \exp\left(-|\Delta s| / k\right)\quad.
\label{eq:PDFdeltas}
\end{equation}
Note that the distribution~(\ref{eq:PDFdeltas}) has zero mean, $\Delta S = \langle \Delta s\rangle = 0$.  The statistical properties of the microadiabatic protocol described here are excellent for an exhaustive experimental study of the fluctuations of a Brownian Carnot engine formed by concatenating isothermal processes with microadabatic protocols.


\section{Brownian Carnot engine \label{sec_brownian} }

\begin{figure}
  \includegraphics[width=\linewidth]{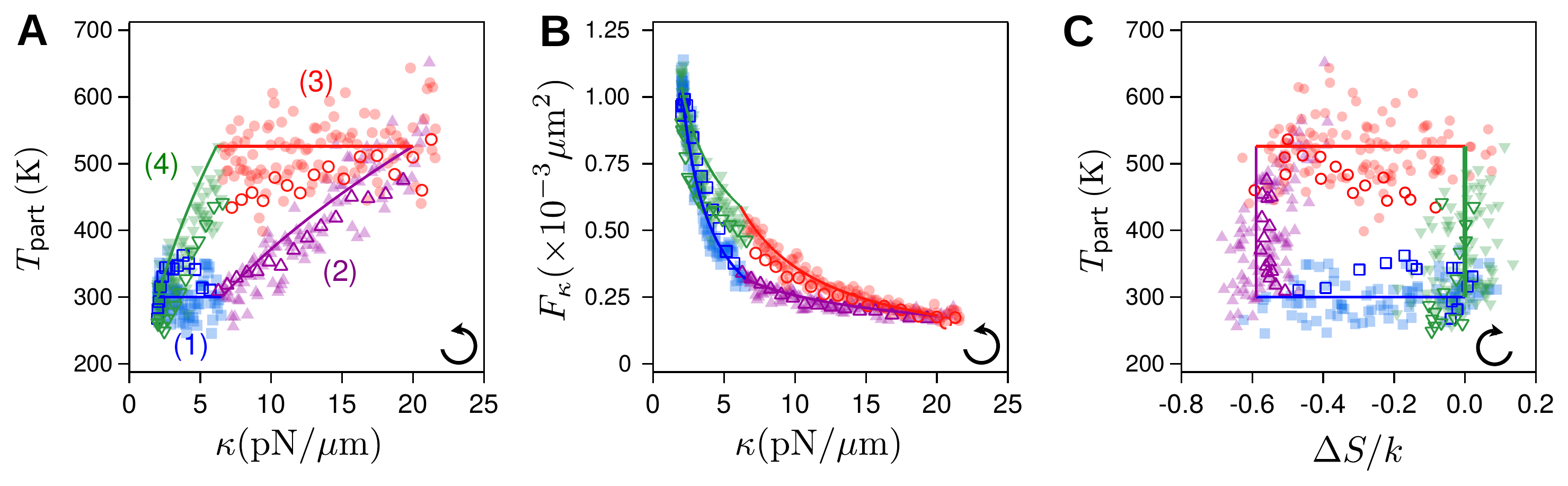}
  \caption{\label{fig_Carnot}  \textbf{(A-C)} \textbf{Thermodynamic diagrams of the engine \cite{martinez2015brownian}: }  (1) Isothermal compression (blue); (2) Adiabatic compression 
 (magenta); (3) Isothermal expansion (red);  (4) Adiabatic expansion (green). Solid lines are the analytical values in the quasistatic limit. Filled symbols are obtained from ensemble averages over cycles of duration $\tau=200\,\rm ms$ while open symbols are obtained for $\tau=30\,\rm ms$. 
The black arrow indicates the direction of the operation of the engine.  \textbf{(A)} $T_{\rm part}-\kappa$ diagram.  
 \textbf{(B)} Clapeyron diagram. The area within the cycle is equal to the mean work obtained during the cycle. \textbf{(C)} $T_{\rm part}$--$S$ diagram. The entropy changes only in the isothermal steps. }
  \centering
\end{figure}

A Carnot cycle is composed of four steps, with hot and a cold isothermal processes joined by two adiabatic steps. Some external parameters must be controlled in a way such that the whole cycle is carried out reversibly, in the present case, the stiffness $\kappa$ and the imposed environment temperature $T$, as shown in figure~\ref{fig_Carnot}A. Both in the cold ($T_c$) and hot ($T_h$) isothermal steps, temperature is held constant (blue and red curves in figure~\ref{fig_Carnot}) whereas $\kappa$ changes in time. As discussed above, for the adiabatic processes, both $\kappa$ and $T$ are varied keeping the ratio $T^2/\kappa$ constant (green and magenta curves in figure~\ref{fig_Carnot}). Figure~\ref{fig_Carnot}B is equivalent to the Clapeyron diagram of the Carnot engine with an ideal gas; the area inside the curve is proportional to the work extracted. The $T_\mathrm{part}-S$ diagram of the particle (figure~\ref{fig_Carnot}C) is a rectangle where all the entropy change occurs in the isothermal steps.
When the stiffness is reduced, the particle probability distribution tends to spread out and we will therefore in the following refer to this as an {\em expansion}. Conversely, an increase in the stiffness is termed a {\em compression}.  

\begin{figure}
  \centering
  \includegraphics[width=0.8\linewidth]{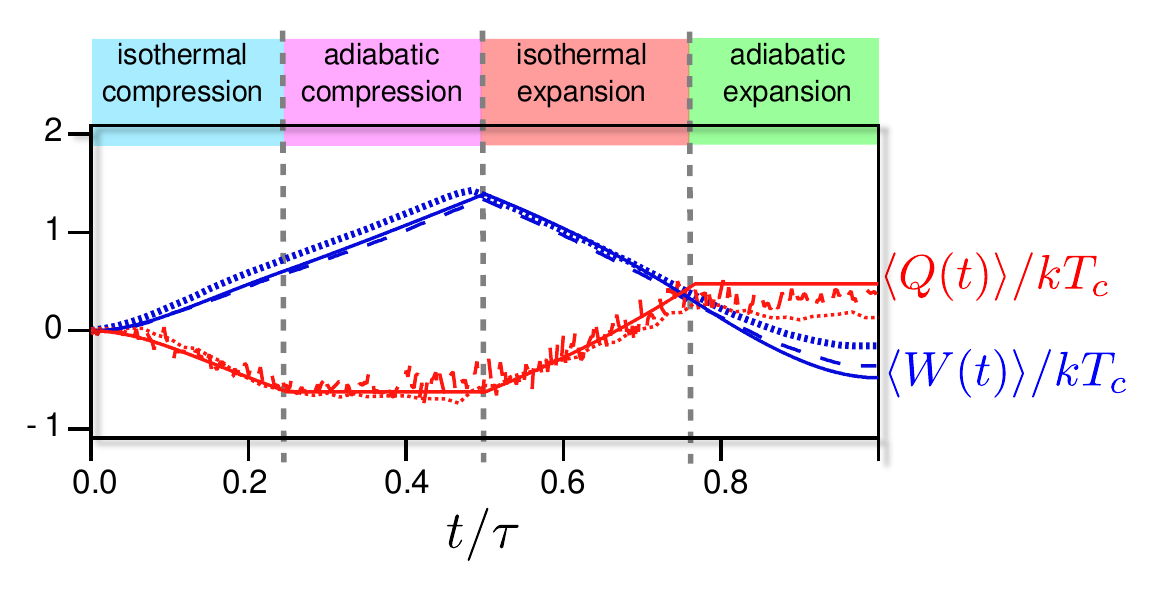}
  \caption{\label{fig_Cumulative_work_and_heat} \textbf{Work and heat in a Carnot cycle.} Accumulated work (blue) and heat (red) exchanged by the system for Carnot cycles of different duration: slow, 200ms (dashed); fast, 30ms (dotted). Continuous line represents the quasistatic theoretical values. During the adiabatic steps heat is not exchanged with the bath, resulting in horizontal lines in accumulated heat. }
  \end{figure}

As in the macroscopic Carnot cycle, work is extracted (on average) in the hot isothermal expansion and a (usually) smaller quantity of work is reinstated to the system in the cold isothermal, making the whole cycle work as an engine, that is, with a total negative work. Work performed in the compressing adiabatic step is compensated (on average) by work extracted in the expanding adiabatic. The total accumulated work and heat along a cycle is represented in figure \ref{fig_Cumulative_work_and_heat}.

\begin{figure}
  \centering
  \includegraphics[width=0.8\linewidth]{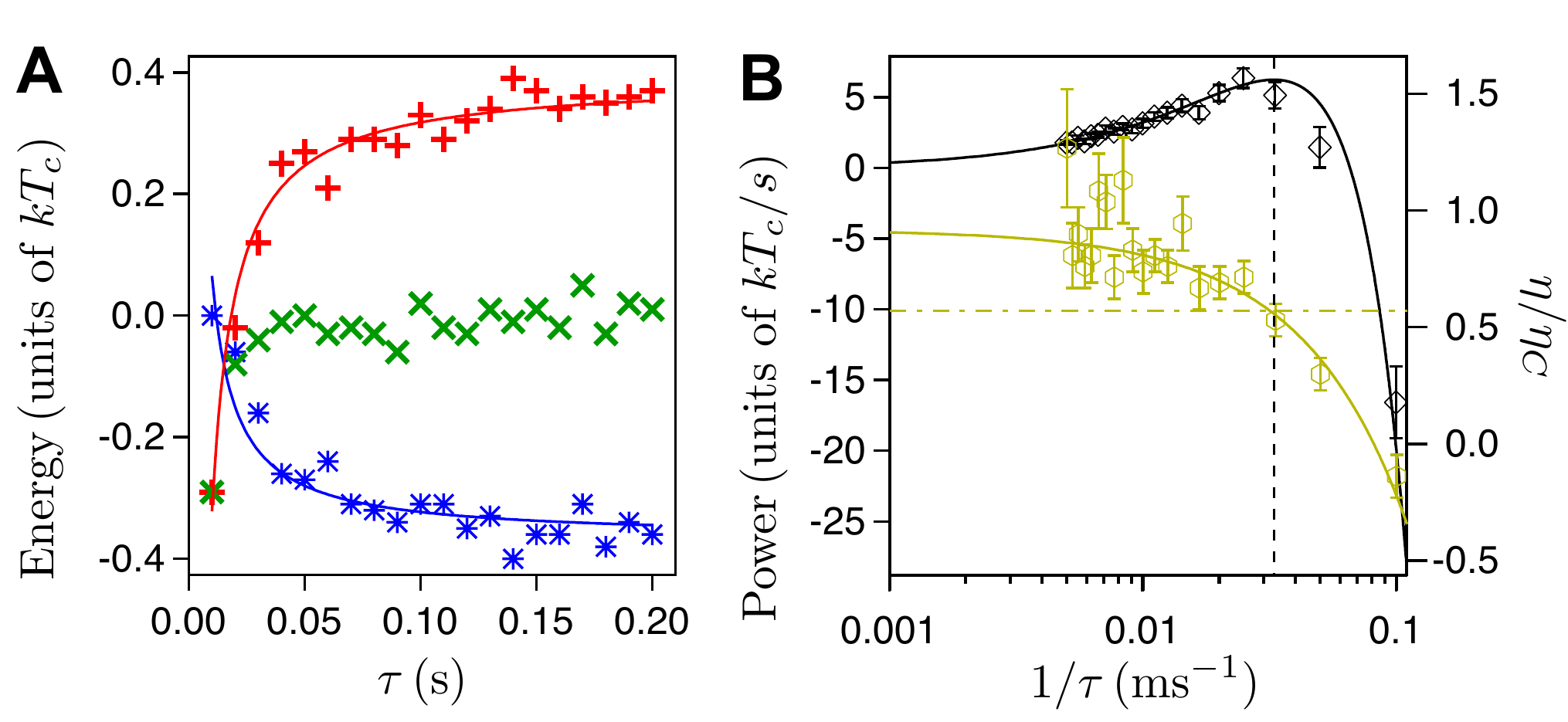}
  \caption{\label{fig_Fig2_Carnot}  {\bfseries Energetics of the Brownian Carnot engine\cite{martinez2015brownian}.} \textbf{(A)} Ensemble averages of stochastic work ($\langle W_\tau\rangle$, blue stars)  and heat ($\langle Q_\tau\rangle$, red pluses) transferred in one cycle as a function of the cycle duration $\tau$. Green crosses are the average total energy change of the working substance $\langle \Delta E_{\mathrm{tot},\tau}\rangle$. Thin lines are fits to Sekimoto-Sasa law, $A+B/\tau$. \textbf{(B)} Power output $P_\tau=-\langle W_\tau\rangle /\tau$  (black diamonds, left axis) and long-term efficiency $\eta_\tau$ (yellow hexagons, right axis) as a function of the inverse of the cycle time. The black  curve is a fit $P_\tau= (\langle W_{\infty}\rangle + \Sigma_{\rm ss}/\tau)/\tau $, yielding  $\langle W_{\infty}\rangle  = (- 0.38\pm 0.01) kT_c$  and $\Sigma_{\rm ss} = (5.7\pm 0.3)kT_c \,\rm ms$ with a reduced chi-square of $\chi^2_{\rm red}=1.08$. The solid yellow line is a fit to $\eta_\tau = (\eta_C+\tau_W/\tau)/(1+\tau_Q/\tau)$, which yields $\eta_{\infty} = (0.92\pm 0.06)\eta_C$, $\tau_W=(-11\pm 2)\,\rm ms$, $\tau_Q=(-0.6\pm 6.0)\,\rm ms$ with $\chi^2_{\rm red}= 0.76$.  Yellow dash-dot line is the Curzon-Alborn efficiency $\eta_{\rm CA}= 1-\sqrt{T_c/T_h} = 0.25=0.57\eta_C$, which is in excellent agreement with the location of the maximum power (vertical black dashed line). Ensemble averages are done over $50\,\rm s$  and error bars are obtained with a statistical significance of $90\%$.  }
\end{figure}

Both work and heat along a cycle of duration $\tau$, $\langle W_\tau\rangle$ and $\langle Q_\tau\rangle$ converge to their quasistatic averages $\langle \cdot_{\infty}\rangle$ following Sekimoto and Sasa's law $\langle W_\tau \rangle =\langle W_{\infty}\rangle + \Sigma_{\rm ss} / \tau$~\cite{sekisasa1997complementarity}, as shown in figure~\ref{fig_Fig2_Carnot}A .  Here, $\langle W_{\infty}\rangle$ is the quasistatic value of the work done per cycle and the term $ \Sigma_{\rm ss} / \tau$ accounts for the (positive) dissipation, which decays to zero like $1/\tau$~\cite{bonanza2014optimal}. We measure the power output as the mean total work exchanged during a cycle divided by the total duration of the cycle (figure~\ref{fig_Fig2_Carnot}B), $P_\tau=-\langle W_\tau\rangle/\tau$.  Except for very fast cycles ($\tau=10\,\rm ms$), the cycle behaves as an engine actually performing work. The power initially increases with $\tau$, and eventually reaches a maximum value  $P_{\rm max} = 6.34\, kT_{\rm c}/s$. Above that maximum, $P_\tau$ decreases monotonically when increasing the cycle length. The data of $P_\tau$ vs $\tau$ fits well to the expected law $P_\tau=-(\langle W_{\infty}\rangle + \Sigma_{\rm ss} / \tau)/\tau$. 

Our engine attains Carnot efficiency when operated quasistatically, that is, when control parameters are changed more slowly than the particle relaxation time. Furthermore, when operated at fast cycle times, it behaves as an endoreversible engine, and its efficiency at maximum power is well described by Curzon and Ahlborn expression~\cite{curzon1975efficiency}. 

The efficiency is given by the ratio between the extracted work and the input of heat, which is usually considered as the heat flowing from the hot thermal bath to the system. Since these quantities fluctuate from cycle to cycle, we can define the following stochastic efficiency:
\begin{equation}
\eta_{\tau}^{1,(i)} = \frac{- W_{\tau}^{(i)}}{  Q_{3,\tau}^{(i)} }\quad
\end{equation}

 where we define $W_{\tau}^{(i)}$ as the sum of the total work exerted on the particle along $i\geq 1$ cycles of duration $\tau$, and $Q_{\alpha,\tau}^{(i)}$ the sum over $i$ cycles of the heat transferred to the particle in the $\alpha-$th subprocess ($\alpha=1,2,3,4$, cf. figure~\ref{fig_Carnot}A).
In our experiment, however, there is a non zero  fluctuating heat in the adiabatic steps, which may be taken into account in the definition of the stochastic efficiency of the engine during a finite number of cycles. Here  we will consider this heat as input, leading to:
\begin{equation}
  \eta_{\tau}^{(i)} = \frac{- W_{\tau}^{(i)}}{ Q_{2,\tau}^{(i)} + Q_{3,\tau}^{(i)}+ Q_{4,\tau}^{(i)}}\quad.
  \label{eq:stoch_eff_3}
\end{equation}
It can be shown that the efficiency $\eta_{\tau}^{(i)}$  defined in (\ref{eq:stoch_eff_3}) has smaller fluctuations than $\eta_{\tau}^{1,(i)}$ \cite{martinez2015brownian} and is therefore more convenient for analysing experimental data.

The long-term efficiency of the motor is given by  $\eta_\tau\equiv\eta_{\tau}^{(i)}$ with $i\to\infty$. In the quasistatic limit, the average heat in the adiabatic processes vanishes and both efficiency definitions coincide, yielding $\eta_\infty =  \eta_{\rm C}\equiv 1-T_{\rm c}/T_{\rm h}\simeq 0.43$ (figure~\ref{fig_Fig2_Carnot}B). 

For fast operation, however, the efficiency decreases, as observed in figure~\ref{fig_Fig2_Carnot}B. This can be understood as originating from an irreversible heat flux between either bath and the particle when in contact at different temperatures.
The environment temperature $T$ controlled by the random noise intensity may differ from the actual effective temperature of the particle obtained from the fluctuations of the position $T_\mathrm{part}=\kappa(t)\langle x(t)^2\rangle/k$, as can be seen in the $T_\mathrm{part}-\kappa$  diagram of the engine (figure~\ref{fig_Carnot}A) or in the $T_\mathrm{part}-S$ diagram (figure~\ref{fig_Carnot}C). 
The standard efficiency at maximum power attained by our engine, $\eta^\star\simeq (0.25\pm 0.05)$, is in agreement with the Curzon-Ahlborn expression for finite-time cycles $\eta_{\rm{CA}}=1-\sqrt{T_c/T_h}\simeq 0.25$~\cite{curzon1975efficiency,esposito2010efficiency}, indicating that the irreversible heat flux is the only irreversibility affecting efficiency at fast operation times.

\begin{figure}
   \centering
   \includegraphics[width=0.8\linewidth]{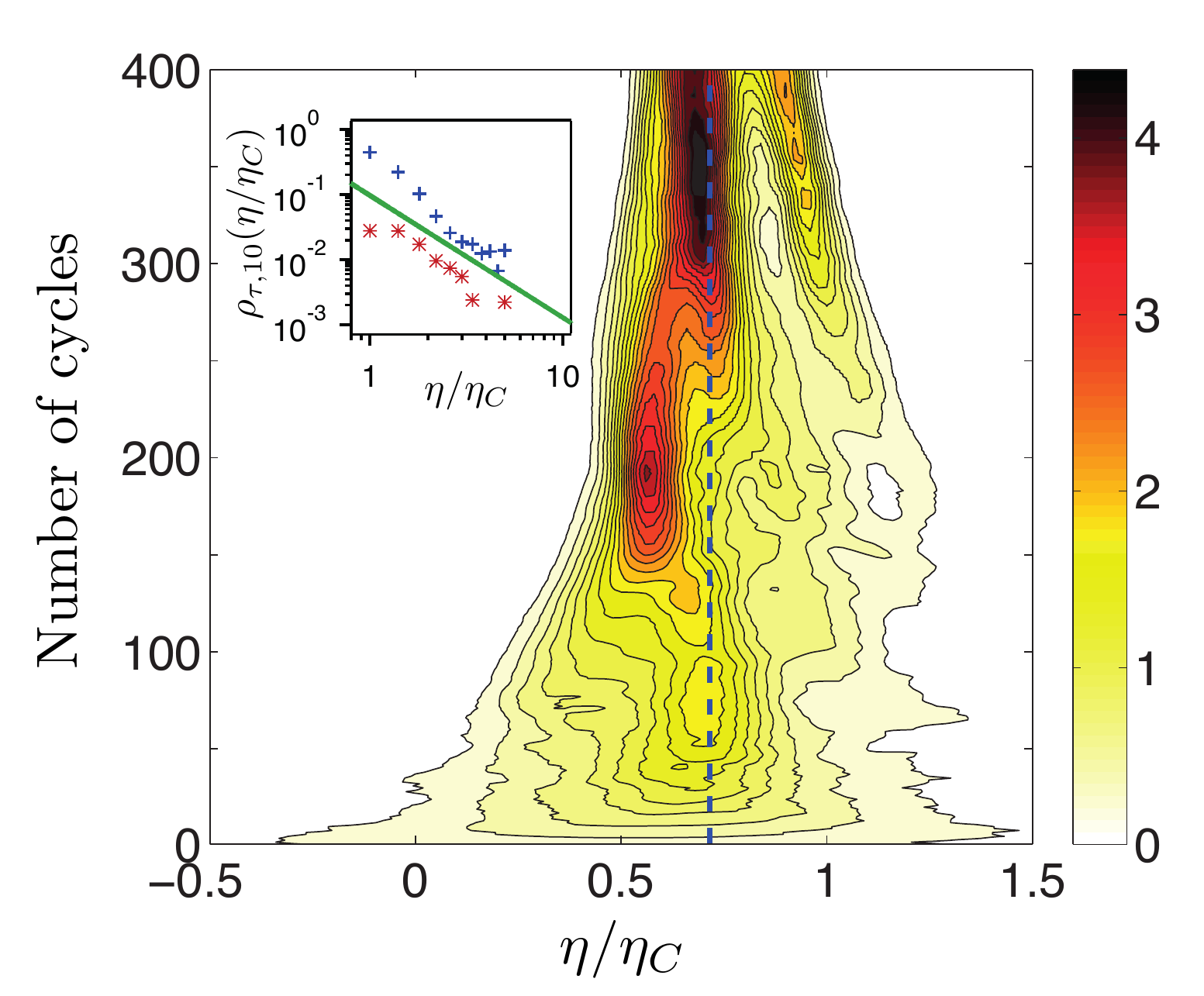}
   \caption{\label{fig_Fig3_Carnot} \textbf{Efficiency fluctuations\cite{martinez2015brownian}:} Contour plot of the probability density function of the efficiency $\rho_{\tau=40\,\mathrm{ms},i} (\eta)$ computed summing over $i=1$ to $400$ cycles (left axis). The  long-term efficiency (averaged over $\tau_{\rm exp} = 50\,\rm s$) is shown with a vertical blue dashed line. Super Carnot efficiencies appear even far from quasistatic driving. {\em Inset}: Tails of the distribution for $\rho_{\tau=40\,\mathrm{ms},10} (\eta)$ (blue squares, positive tail; red circles, negative tail). The green line is a fit to a power-law to all the data shown, whose exponent is $\gamma = (-1.9\pm 0.3)$. }
 \end{figure}

Finally, we used our setup to test the two main theoretical predictions about the probability distribution of the fluctuations or the stochastic efficiency. In particular, sufficiently close to equilibrium so that fluctuations of work and heat can be considered Gaussian, near the maximum power output of the engine, the distribution is bimodal when summing over several cycles (figure~\ref{fig_Fig3_Carnot})~\cite{gingrich2014efficiency,polettini2015efficiency}.  Indeed, local maxima of $\rho_{\tau,i}(\eta)$ appear above standard efficiency for large values of $i$. Another universal feature tested is that the tails of the distribution follow a power-law, $\rho_{\tau,i}(\eta\to\pm\infty)\sim\eta^{-2}$ (inset of figure~\ref{fig_Fig3_Carnot})~\cite{gingrich2014efficiency,proesmans2015stochastic}.  
The stochastic nature of the efficiency rises a number of questions that will be discussed in section \ref{sec_open_problems}.

\section{Conclusion and open problems \label{sec_open_problems}}

\begin{figure}
  \centering
  \includegraphics[width=8cm]{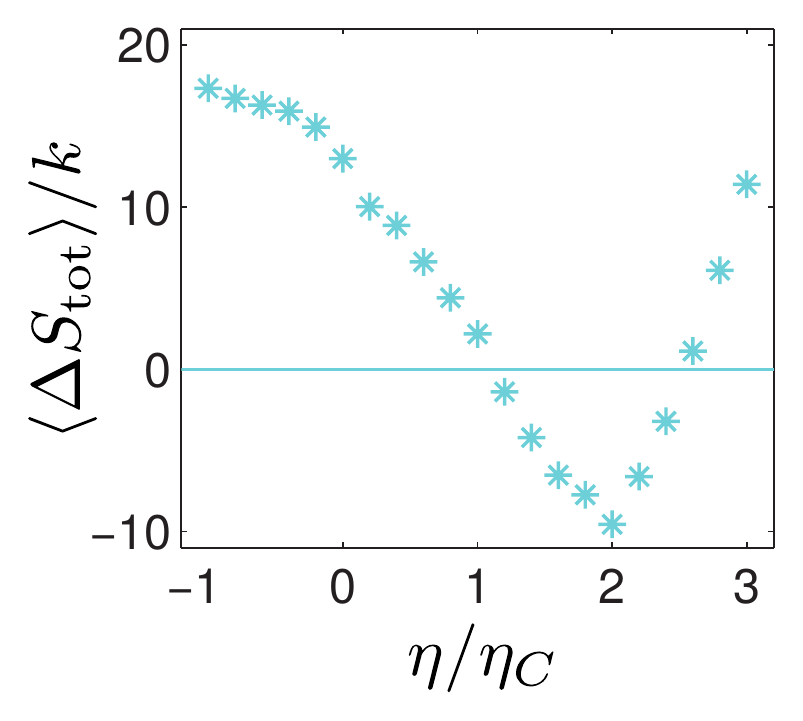}
  \caption{\label{fig_LDF_and_entropy_s2} \textbf{Mean entropy production \cite{martinez2015brownian}} as a function of the efficiency for an observation time of $30$ cycles.  Mean entropy production vanishes at $\eta\simeq \eta_C$ and near $\eta\simeq \eta_{\rm min}$. Here $\eta\simeq \eta_{\rm min}>\eta_C$ represents a value of the efficiency where the PDF of effiency has a relative minimum.}
\end{figure}

We have demonstrated how to construct a Carnot engine with a single microscopic particle as a working substance which is able to transform the heat transferred from thermal fluctuations into mechanical work. At slow driving, our engine attains the fundamental limit of Carnot efficiency.  There are however a number of differences between a macroscopic Carnot engine and its Brownian analogue. At these scales, energy exchanges become stochastic. As a consequence, efficiency for instance becomes fluctuating allowing for particular realizations with Carnot or super-Carnot efficiency even far from equilibrium. {Furthermore, it remains also an open question whether the definition of (stochastic) efficiency should necessarily include the heat in other steps than the hot isothermal or not}.  Finally, in contrast to the macroscopic counterpart, the microscopic Carnot engine displays reversible trajectories at finite power, where the average entropy production vanishes. Figure~\ref{fig_LDF_and_entropy_s2}  shows that entropy production may vanish not only for efficiency equal to Carnot value but also above this, a consequence of the bimodality of the efficiency PDF. Whether one can profit from finite power reversible trajectories or fluctuations in general in motors operating between two thermal reservoirs remains an open question. It may be possible to use better design principles in order to enhance the efficiency of fluctuating thermal motors \cite{2015arXiv150305788M,PhysRevLett.106.250601}
 by keeping them working close to the vanishing entropy production regime at finite power.

Finally, it may be worth analysing the possibility of applying the techniques for adiabatic steps described here to other microscopic systems. Isolation may be of interest in itself in micro or nano devices operating in contact with several heat baths. 

\section{Acknowledgements}
{L.D., E.R. and J.M.R.P. acknowledge financial support from Spanish Government, grants ENFASIS (FIS2011-22644) and TerMic (FIS2014-52486-R). IAM acknowledges financial support from European  Research  Council grant OUTEFLUCOP.  This work is partly supported by the German Research Foundation (DFG) within the Cluster of Excellence 'Center for Advancing Electronics Dresden'. 
  R.A.R. acknowledges financial support from Fundació Privada Cellex and  from the Ministerio de Economía y Competitividad (MINECO), Spain, through project NanoMQ (FIS2011-24409) and Severo Ochoa Excellence Grant (SEV-2015-0522).
D. Petrov, leader of the Optical Tweezers group at ICFO,
who passed away on 3rd February 2014. Fruitful discussions with P. Mestres and A. Ortiz-Ambriz are gratefully acknowledged. }   

\appendix
\section{Energetics of quasistatic processes \label{sec_appendix_quasistatic}}

 The work and heat along a quasistatic process of total duration $\tau$ averaged over many realizations are equal to
 \begin{equation}
   \langle W \rangle = \int_{0}^{\tau} \frac{ \langle x^2(t)\rangle}{2} d\kappa(t)  = 
   \int_{0}^{\tau}   \frac{kT(t)}{2\kappa(t)}\,d\kappa(t),
   \label{work}
 \end{equation}
 \begin{equation}
   \langle Q \rangle = \int_{0}^{\tau} \frac{\kappa(t)}{2}\, d\left[\langle x^2(t)\rangle\right]  = 
   \int_{0}^{\tau}  \frac{\kappa(t)}{2}\, d\left(\frac{kT(t)}{\kappa(t)}\right),
   \label{heatX}
 \end{equation}
 where we have used  equipartition theorem along the quasistatic protocols, $\kappa(t)\langle x^2 (t)\rangle  = k T (t)$.  The average potential and kinetic energy changes are equal to  $\langle \Delta U \rangle= \langle \Delta E_{\rm kin} \rangle=  \frac{k}{2} [T(\tau)-T(0)]$. Finally, the total energy change is $\langle \Delta E_{\rm tot} \rangle = \langle \Delta U \rangle + \langle \Delta E_{\rm kin} \rangle=  k [T(\tau)-T(0)]$. Therefore, for any protocol where $\kappa$ and $T$ are changed in a controlled way, all the values of the energy exchanges are known, and can therefore be compared with measurements.

\bibliography{refs_UPoN15}{}
\bibliographystyle{unsrt}

\end{document}